\documentclass[prd,superscriptaddress,amsfonts,amssymb,amsmath,showpacs,twocolumn,nofootinbib]{revtex4-2}
\usepackage{bm}
\usepackage{amsfonts}
\usepackage{latexsym}
\usepackage{graphicx}
\usepackage{amsmath}
\usepackage{palatino}
\usepackage{xcolor} 
\usepackage{mathpazo}
\usepackage{xcolor}
\usepackage{rotating}
\usepackage{adjustbox}
\usepackage{tensor}
\usepackage{textcomp}
\usepackage{float}
\usepackage{booktabs}
\usepackage{dcolumn}
\usepackage[titletoc]{appendix}
\usepackage{booktabs}
\usepackage{multirow}
\usepackage{hyperref}
\hypersetup{colorlinks,citecolor=blue}
\usepackage{amsmath}
\usepackage{xcolor}
\usepackage{orcidlink}
\usepackage{epsfig}
\usepackage{caption}
\usepackage{subcaption}
\usepackage{commath}
\usepackage[normalem]{ulem}

\hypersetup{colorlinks,citecolor=blue}
\hypersetup{colorlinks=true,linkcolor=magenta,filecolor=magenta,    urlcolor=blue}

\def\be{\begin{equation}}
\def\ee{\end{equation}}
\def\bea{\begin{eqnarray}}
\def\eea{\end{eqnarray}}

\begin{document}

\title{No preference for generalized emergent dark energy from current cosmological data}

\author{Vipin Kumar Sharma} 
\email{vipin.33912@lpu.co.in\\ vipinastrophysics@gmail.com}
\affiliation{International Center for High Energy Physics and Applications, Lovely Professional University, Phagwara, Punjab, 144411, India.}
\affiliation{School of Chemical Engineering and Physical Sciences, Lovely Professional University, Phagwara, Punjab, 144411, India}

\author{Himanshu Chaudhary}
\email{himanshu.chaudhary@ubbcluj.ro,\\
himanshuch1729@gmail.com}
\affiliation{Department of Physics, Babeș-Bolyai University, Kogălniceanu Street, Cluj-Napoca, 400084, Romania}
\affiliation{Research Center of Astrophysics and Cosmology, Khazar University, Baku, AZ1096, 41 Mehseti Street, Azerbaijan}
\author{Salvatore Capozziello}
\email{capozziello@na.infn.it}
\affiliation{Dipartimento di Fisica ``E. Pancini", Universit\`a di Napoli ``Federico II", Complesso Universitario di Monte Sant’ Angelo, Edificio G, Via Cinthia, I-80126, Napoli, Italy,}
\affiliation{Istituto Nazionale di Fisica Nucleare (INFN), sez. di Napoli, Via Cinthia 9, I-80126 Napoli, Italy,}
\affiliation{Scuola Superiore Meridionale, Via Mezzocannone 4, I-80134, Napoli, Italy.} 
\author{Dhruba Jyoti Gogoi}
\email{moloydhruba@yahoo.in}
\affiliation{Department of Physics, Madhabdev University, Narayanpur, Lakhimpur 784164, Assam, India}
\author{G. Mustafa}
\email{gmustafa3828@gmail.com}
\affiliation{Department of Physics,
Zhejiang Normal University, Jinhua 321004, People’s Republic of China}
\affiliation{College of Graduate Studies, Walailak University, Nakhon Si Thammarat, Thailand}
\affiliation{Zhejiang Institute of Photoelectronics and Zhejiang Institute for Advanced Light Source, Jinhua 321004, China PDU}

\begin{abstract}
In this work, we revisit the generalized emergent dark energy model by confronting it with DESI DR2 baryon acoustic oscillation measurements, in combination with joint CMB data from ACT, SPT, and Planck, as well as Type Ia supernova samples including Pantheon$^+$, DES-Dovekie, and Union3. We find that the GEDE model remains compatible with the $\Lambda$CDM paradigm, with no statistically significant preference for deviations when all datasets are combined. In particular, the key model parameter $\Delta$ is consistent with the $\Lambda$CDM value $\Delta = 0$ within $2\sigma$ once SNe Ia data are included. Despite this overall agreement, the GEDE model does not exhibit the phantom crossing suggested by DESI DR2. Instead, the evolution of the dark energy equation of state $w(z)$ indicates that the model behaves either as a full phantom ($w < -1$) or quintessence ($w > -1$), depending on the dataset combination, without a clear transition across $w = -1$. When Pantheon$^+$ data are included, the model converges closely to $\Lambda$CDM with $w \simeq -1$. The GEDE model does not alleviate the existing cosmological tensions. The inferred values of $H_0$ remain in the range $67.9$–$69.8\ \mathrm{km,s^{-1},Mpc^{-1}}$, while the sound horizon $r_d$ and clustering parameter $S_8$ remain consistent with $\Lambda$CDM, failing to resolve the $H_0$ and $S_8$ tensions. Finally, both Gaussian significance levels ($<2\sigma$) and Bayesian evidence ($\ln B_{i,j} < 1$) indicate no statistically significant preference for GEDE over $\Lambda$CDM. We conclude that, although DESI DR2 hints at dynamical dark energy, the GEDE model remains observationally indistinguishable from $\Lambda$CDM and does not support the Quintom-B–type behavior suggested by DESI DR2. 
\end{abstract}

\maketitle

\section{Introduction}\label{sec_1}
The release of the Dark Energy Spectroscopic Instrument Data Release 2 has intensified the ongoing debate about the adequacy of the $\Lambda$CDM paradigm in describing the late-time Universe \cite{abdul2025desi,popovic2025dark,Zhao:2017cud,Cortes:2024lgw,Capozziello:2025qmh,DESI:2024aqx}. Although the six defining parameters of flat $\Lambda$ Cold Dark Matter ($\Lambda$CDM) remains the cornerstone of modern cosmology, its reliance on a cosmological constant ($\Lambda$) with a fixed equation-of-state parameter $w = -1$ has long been subject to challenges on both theoretical and observational grounds \cite{Nesseris:2006er,betoule2014improved,aghanim2020planck1,aghanim2020planck2,Wolf:2023uno,efstathiou1999cosmic,kosowsky2002efficient,Wang:2018fng}. The cosmological constant problem, the coincidence problem, and persistent tensions in key 
parameters such as $H_{0}$ and $S_{8}$ have motivated the exploration of dynamical dark energy scenarios \cite{Sahni:1999gb,Padmanabhan:2002ji,Bagla:1995az,DiValentino:2020zio,Pandey:2019plg,Bhattacharyya:2018fwb,DiValentino:2021izs,DiValentino:2020vvd}. DESI DR2 offers highly precise BAO measurements across many galaxy types and redshift ranges, making it an important way to test models beyond $\Lambda$CDM (phenomenological Dark Energy,  and or Modified Gravity) \cite{Chevallier:2000qy,Linder:2002et,Jassal:2004ej,Joyce:2016vqv}.

Recently, we analysed several possible alternatives to the standard concordance cosmological model. The DESI DR2 results show small to moderate mismatches with $\Lambda$CDM predictions, especially in the matter density parameter $\Omega_{m}$ measured from luminous red galaxies (LRGs) and emission line galaxies (ELGs) \cite{Chaudhary:2025pcc,Chaudhary:2025uzr,Chaudhary:2025bfs,Capozziello:2025lor}. These discrepancies, though 
not yet decisive, suggest that dark energy may evolve with redshift rather than remain constant. Several studies have interpreted the preference for $w_{0} > -1$ and $w_{a} < 0$ within the Chevallier–Polarski–Linder (CPL) framework as indicative of a Quintom-B type scenario, where the dark energy equation of state crosses the cosmological constant boundary \cite{Chaudhary:2025bfs,Cai:2025mas,Malekjani:2025alf}. In parallel, researchers have investigated generalized emergent dark energy models \cite{Yang:2021eud,Li:2020ybr,Sharma:2025qmv}, which posit that dark energy was negligible in the early Universe and only manifested dynamically at late times, thereby offering a natural resolution to the coincidence problem. 

The generalized emergent dark energy framework  is particularly compelling in light of DESI DR2. Analyses  combining BAO with Type Ia supernovae (Pantheon$^+$, Union3, DES-SN5Y) and cosmic chronometers consistently show deviations from $\Lambda$CDM at the $2\sigma$–$4\sigma$ level, with Bayesian model comparison yielding weak-to-moderate evidence in favor of dynamical or emergent scenarios \cite{Sharma:2025qmv}. Notably, GEDE retains the simplicity of $\Lambda$CDM while introducing a single free parameter that governs the onset of dark energy dominance, allowing it to interpolate between $\Lambda$CDM  ($\Delta = 0$) and PEDE ($\Delta = 1$) limits. In our previous study, the constraints from DESI DR2 suggest that GEDE can fit BAO and supernova data comparably to $\Lambda$CDM, while offering improved consistency at intermediate redshifts where $\Lambda$CDM tensions are most pronounced.

This study explores the theoretical framework  of GEDE, in which we aim to critically assess whether DESI DR2 provides confirmatory evidence for generalised emergent dark energy. We situate GEDE within the broader landscape of dynamical dark energy models including CPL, logarithmic, exponential, and Quintom frameworks and assess its performance against DESI DR2 BAO measurements, supernova compilations, and a full CMB analysis, which had not been comprehensively considered in earlier studies. By employing nested sampling and Bayesian evidence metrics, we investigate whether generalised emergent dark energy is  merely a phenomenological extension or a genuinely preferred description of the late-time Universe. In doing so, we address the central question: does DESI DR2 mark the beginning of a paradigm shift away from $\Lambda$CDM toward generalised emergent dark energy? 

The paper is structured as follows. Section~\ref{sec_2} outlines the governing equations of the GEDE model. Section~\ref{sec_3} presents the methodology employed to constrain the model’s free parameters and provides an overview of the observational datasets utilized. Section~\ref{sec_4} reports the results of our analysis and concludes with a discussion of our findings and their broader implications.
 
\section{Theoretical background}\label{sec_2}

We consider a homogeneous and isotropic Universe well approximated by the spatially flat Friedmann–Lemaître–Robertson–Walker (FLRW) metric. The expansion history is governed by the Friedmann equations, which include contributions from radiation, matter, and dark energy. The first Friedmann equation reads
\begin{equation}
H^2(z) = \frac{8\pi G}{3}\left[\rho_{r}(z) + \rho_{m}(z) + \rho_{de}(z)\right],\label{1}
\end{equation}
where $H(z)$ is the Hubble expansion rate, $\rho_{r}$ is the radiation density, $\rho_{m}$ is the matter density, and $\rho_{de}$ is the dark energy density. The continuity equation for each component is
\begin{equation}
\dot{\rho}_{x} + 3H(1+w_{x})\rho_{x} = 0,\label{2}
\end{equation}
with $w_{x}$ the equation-of-state (EoS) parameter for component $x$ (radiation, matter, or dark energy).

\subsection{The $\Lambda$CDM Model}
In the concordance $\Lambda$CDM model, dark energy is described by a cosmological constant 
with $w_{de} = -1$. The dark energy density remains constant, $\rho_{de}(z) = \rho_{de,0}$, 
and the dimensionless Hubble parameter is

\begin{equation}
\begin{split}
E^2(z) \equiv \frac{H^2(z)}{H_0^2} & = \Omega_{r}(1+z)^4 +\Omega_{m}(1+z)^3 + \Omega_{\mathrm{DE}} \frac{\rho_{\mathrm{DE}}(z)}{\rho_{\mathrm{DE},0}} , \label{3}
\end{split}
\end{equation}
where  $\Omega_{r}$, $\Omega_{m}$, and $\Omega_{\mathrm{DE}}$ denote the energy densities in radiation, matter, and dark energy, respectively. Here, $H_0$ represents the Hubble constant. In our analysis, we assume a spatially flat Universe.

\subsection{Genralized Emergent Dark Energy framework}

An alternative approach is the emergent dark energy scenario, where dark energy is negligible in the early Universe and becomes dominant only at late times \cite{Yang:2021eud}. The Generalized Emergent Dark Energy (GEDE) model introduces a dimensionless parameter $\Delta$ controlling the emergence of dark energy \cite{Li:2020ybr}. Here the normalized dark energy density evolves as
\begin{equation}
f_{DE}(z) = \frac{1 - \tanh\left[\Delta \log_{10}\left(\frac{1+z}{1+z_t}\right)\right]}
{1 + \tanh\left[\Delta \log_{10}(1+z_t)\right]},\label{7}
\end{equation}
with $z_t$ as the transition redshift at which $\rho_{m}(z_t) = \rho_{de}(z_t)$. Considering this, the dimensionless Friedmann equation, expressed in terms of normalized density parameters, is given by
\begin{equation}\label{eq_110}
\begin{split}
    E^2(z) = & \, \Omega_{r}(1+z)^4 +  \Omega_{m}(1+z)^3 \\
    & + \Omega_{\text{DE},0} \left( \frac{1 - \tanh\!\left(\Delta \cdot \log_{10} \!\left( \frac{1+z}{1+z_t} \right)\right)}{1 + \tanh\!\left(\Delta \cdot \log_{10} (1 + z_t)\right)} \right),
\end{split}
\end{equation}

The parameter \(\Delta\) is dimensionless and free, with notable characteristics: setting \(\Delta = 0\) reduces equation \eqref{eq_110} to the standard \(\Lambda\)CDM model with constant \(w = -1\).

The equation of state for the GEDE model can be obtained from equation \eqref{7} using
\begin{equation}\label{eq_60}
w(z) = \frac{1}{3}(1+z)\frac{d\ln \tilde{\Omega}_{DE}}{dz} - 1,
\end{equation}
which yields
\begin{equation}\label{eq_601}
w(z) = -1 - \frac{\Delta}{3\ln (10)} \left[ 1 + \tanh\!\left( \Delta \cdot \log_{10}\!\left(\frac{1+z}{1+z_t}\right)\right)\right].
\end{equation}
We now confront the extended Friedmann equations with current cosmological observations, performing a full CMB analysis alongside BAO and supernova datasets in next section.

\section{Dataset and Methodology}\label{sec_3}

To constrain the parameters of the cosmological models, we use the {\tt Cobaya}\footnote{\url{https://github.com/CobayaSampler/cobaya}} \cite{torrado2021cobaya}, a code for bayesian inference that allows us to explore the arbitrary priors and posteriors using a range of monte carlo samplers. These include the Markov Chain Monte Carlo (MCMC) sampler from CosmoMC and the nested sampling algorithm {\tt PolyChord} \cite{handley2015polychord1,handley2015polychord2}. In this analysis, we employ the MCMC algorithm \cite{lewis2002cosmological,lewis2013efficient,neal2005taking}. The convergence of the chains is assessed using the Gelman–Rubin criterion \cite{gelman1992inference}, requiring $R - 1 < 0.01$. The theoretical model are computed using the boltzmann solver Code for Anisotropies in the Microwave Background {\tt (CAMB)}\footnote{\url{https://camb.readthedocs.io/en/latest/}} code \cite{lewis2000efficient,howlett2012cmb}. After convergence, the sampling is analysed using the \texttt{GetDist} package \cite{lewis2025getdist}. During our analysis, for the $\Lambda$CDM model, the set of free parameters is given by

$\bm{\theta}_{\Lambda\mathrm{CDM}} = \{\Omega_{\mathrm{cdm}}, \Omega_{\mathrm{b}}, 100\,\theta_{\mathrm{MC}}, \ln(10^{10}A_{\rm s}), n_{\rm s}, \tau \}$. For the GEDE model, the corresponding parameter set is extended to $\bm{\theta}_{\mathrm{GEDE}} = \{\bm{\theta}_{\Lambda\mathrm{CDM}}, \Delta\}.$

We also use the Bayesian evidence selection criterion to identify the preferred GEDE model over the $\Lambda$CDM model. The Bayesian evidence, $\ln \mathcal{Z}$, is computed using the \texttt{MCEvidence} \cite{heavens2017marginal,heavens2017no}, through the Cobaya interface available in the \texttt{wgcosmo} repository\footnote{\url{https://github.com/williamgiare/wgcosmo.git}}. The Bayesian evidence $\mathcal{Z}$ is defined as the integral of the likelihood function over the parameter space, $\mathcal{Z} = \int_{\Omega} P(D \mid \boldsymbol{\theta}, M)\,
P(\boldsymbol{\theta} \mid M)\, P(M)\, d\boldsymbol{\theta},$ where $P(D \mid \boldsymbol{\theta}, M)$ denotes the likelihood of the observational data $D$ given the model parameters $\boldsymbol{\theta}$, $P(\boldsymbol{\theta} \mid M)$ represents the prior distribution of the parameters, $P(M)$ is the prior probability assigned to the model, and $\Omega$ corresponds to the allowed parameter space. The relative preference between the two models is quantified using the Bayes factor, evaluated in logarithmic form as $\ln B = \ln \mathcal{Z}_{\mathrm{GEDE}} - \ln \mathcal{Z}_{\Lambda\mathrm{CDM}},$ where $\mathcal{Z}_{\mathrm{GEDE}}$ and $\mathcal{Z}_{\Lambda\mathrm{CDM}}$ are the Bayesian evidences of the GEDE and $\Lambda$CDM models, respectively.

The statistical significance of the model comparison is interpreted according to the Jeffreys scale \cite{kass1995bayes,trotta2008bayes}. Within this framework, values of  $\ln B_{\mathrm{GEDE},\Lambda\mathrm{CDM}} < 1$ correspond to inconclusive evidence; $1 \leq \ln B_{\mathrm{GEDE},\Lambda\mathrm{CDM}} < 2.5$ indicate weak evidence; $2.5 \leq \ln B_{\mathrm{GEDE},\Lambda\mathrm{CDM}} < 5$ are classified as moderate evidence; $5 \leq \ln B_{\mathrm{GEDE},\Lambda\mathrm{CDM}} < 10$ signify strong evidence; and $\ln B_{\mathrm{GEDE},\Lambda\mathrm{CDM}} \geq 10$ provide decisive support in favor of the GEDE model.

To constrain the parameter of the GEDE model, we use several observational data from Baryon Acoustic Oscillations, Type~Ia Supernovae, and Cosmic Microwave Background likelihoods, which are described in detail below.

\begin{itemize}
     \item \textbf{Baryon Acoustic Oscillation :} First, we use the Baryon Acoustic Oscillation (BAO) measurements from the Dark Energy Spectroscopic Instrument (DESI) Data Release~2\footnote{\url{https://github.com/CobayaSampler/bao_data.git}} (DR2) \cite{abdul2025desi}, extracted from multiple tracers including the Bright Galaxy Sample (BGS), Luminous Red Galaxies (LRG1-3), Emission Line Galaxies (ELG1-2), Quasars (QSO), and the Lyman-$\alpha$ forest. These measurements span seven redshift bins in the range $0.3 \leq z \leq 2.33$. The DESI DR2 BAO information is incorporated through the compressed distance measures $D_M/r_d$, $D_H/r_d$, and $D_V/r_d$, where $D_H(z) = c/H(z)$ is the Hubble distance, $D_M(z) = c \int_0^z dz'/H(z')$ is the transverse comoving distance, $D_V(z) = [z D_M^2(z) D_H(z)]^{1/3}$ is the isotropic distance, and $r_d$ denotes the sound horizon at the baryon drag epoch.
     \item \textbf{Type Ia Supernova :} Then, we also use three Type~Ia supernova (SNe~Ia) catalogs. The first is the Pantheon$^+$\footnote{\url{https://github.com/PantheonPlusSH0ES/DataRelease.git}} compilation \cite{scolnic2022pantheon}, which contains 1{,}701 light curves from 1{,}550 SNe~Ia over the redshift range $0.001 \leq z \leq 2.26$. To mitigate systematic effects from peculiar velocities at very low redshift, we exclude light curves with $z < 0.01$, leaving a final sample of 1{,}590 light curves \cite{brout2022pantheon}. Then, we use the recalibrated Dark Energy Survey (DES)-Dovekie\footnote{\url{https://github.com/des-science/DES-SN5YR.git}} SNe~Ia sample \cite{popovic2025dark}, which consists of 1{,}820 photometric light curves over the redshift range of $0.02 \leq z \leq 1.14$. Among these, 1{,}623 SNe~Ia were observed by DES, while 197 correspond to low-redshift SNe~Ia from external surveys such as CfA and CSP \cite{hicken2009cfa3,hicken2012cfa4,foley2017foundation}. The revised DES-Dovekie sample shares 1{,}718 SNe~Ia with the DES 5-year Type Ia supernova sample (DES-SN5YR) \cite{abbott2024dark,sanchez2024dark,vincenzi2024dark}. Finally, we use the Union3\footnote{\url{https://github.com/rubind/union3_release}} compilation \cite{rubin2025union}, consisting of 2{,}087 SNe~Ia over the redshift range $0.05 < z < 2.26$, of which 1{,}363 are common to the Pantheon$+$ sample.
     \item \textbf{CMB likelihood:} Finally, we use the following combination of CMB datasets: 1) We use the \texttt{CamSpec} CMB likelihood \cite{Efstathiou2021Detailed,rosenberg2022cmb} built from the latest Planck \texttt{NPIPE PR4} data release from the Planck collaboration \cite{aghanim2020planck1,aghanim2020planck2}, including the temperature (TT), polarization (EE), and cross-correlation (TE) power spectra from Planck, specifically using the \texttt{SimAll}, \texttt{Commander} ($\ell < 30$), and \texttt{CamSpec} ($\ell \geq 30$) likelihoods. 2) The SPT-3G main field data, including temperature and polarization (TT/TE/EE) anisotropy spectra, is implemented using the SPT-lite likelihood\footnote{\url{https://github.com/SouthPoleTelescope/spt_candl_data}.} \cite{camphuis2025spt}. 3) The ACT DR6 CMB likelihood, providing small-scale temperature and polarization power spectra (TT: $1000 \leq \ell \leq 8500$; TE/EE: $600 \leq \ell \leq 8500$), is implemented using the \texttt{ACT-lite} likelihood \footnote{\url{https://github.com/ACTCollaboration/DR6-ACT-lite}}. 4) The combination of the ACT DR6, SPT-3G, and Planck NPIPE lensing likelihoods \cite{maccrann2024atacama,qu2024atacama,madhavacheril2024atacama,qu2026unified,ge2025cosmology} \footnote{\url{https://github.com/qujia7/spt_act_likelihood.git}}.
\end{itemize}
The priors chosen for the $\Lambda$CDM and GEDE models are summarized in Table~\ref{tab_1}.

{
\renewcommand{\arraystretch}{1}
\begin{table}[t] 
    \centering
    \begin{tabular}{|lll|}
    \hline
    parametrization & parameter & prior\\  
    \hline 
    $\mathbf{\Lambda}$\textbf{CDM} & $\Omega_\mathrm{cdm}h^2$ & $\mathcal{U}[0.001, 0.99]$ \\   
    & $\Omega_\mathrm{b}h^{2}$ & $\mathcal{U}[0.005, 0.1]$ \\
    & $100\theta_\mathrm{MC}$ & $\mathcal{U}[0.5, 10]$ \\
    & $n_\mathrm{s}$ & $\mathcal{U}[0.8, 1.2]$ \\
    & $\tau_{\mathrm{reio}}$ & $\mathcal{U}[0.01, 0.8]$ \\
    & $\ln(10^{10} A_\mathrm{s})$ & $\mathcal{U}[1.61, 3.91]$ \\
    \hline 
    \textbf{GEDE} & $\Delta$ & $\mathcal{U}[-3, 10]$ \\
    \hline
    \end{tabular}
    \caption{ Parameters and priors used in the analysis. Here $\mathcal{U}[{\rm min, max}]$ denotes a uniform prior over the specified range.}\label{tab_1}
\end{table}
}

\begin{table*}[htbp]
\centering
\resizebox{\textwidth}{!}{%
\begin{tabular}{l@{\hspace{10pt}}c@{\hspace{10pt}}c@{\hspace{10pt}}c@{\hspace{10pt}}c}
\hline
\textbf{Parameters} 
& \textbf{CMB + DESI DR2} 
& \textbf{CMB + DESI DR2 + Pantheon$^+$} 
& \textbf{CMB + DESI DR2 + DES-Dovekie} 
& \textbf{CMB + DESI DR2 + Union3} \\
\hline

$\Omega_{\mathrm{\text{cdm}}}h^2$ 
& $0.11878 \pm 0.00075$ 
& $0.11808 \pm 0.00069$
& $0.11799 \pm 0.00071$
& $0.11812 \pm 0.00073$ \\

$\Omega_{\mathrm{\text{b}}}h^2$ 
& $0.022412 \pm 0.000084$ 
& $0.022437 \pm 0.000083$
& $0.022436 \pm 0.000085$
& $0.022436 \pm 0.000084$ \\

$100\,\theta_{\mathrm{MC}}$ 
& $1.04094 \pm 0.00018$ 
& $1.04100 \pm 0.00017$
& $1.04101 \pm 0.00017$
& $1.04099 \pm 0.00017$ \\

$\tau_{\mathrm{reio}}$ 
& $0.0624_{-0.0078}^{+0.0070}$ 
& $0.0666_{-0.0082}^{+0.0068}$
& $0.0671_{-0.0081}^{+0.0071}$
& $0.0665_{-0.0082}^{+0.0074}$ \\

$n_{\mathrm{s}}$ 
& $0.9694 \pm 0.0030$ 
& $0.9710 \pm 0.0030$
& $0.9711 \pm 0.0029$
& $0.9708 \pm 0.0030$ \\

$\ln\bigl(10^{10}\,A_{\mathrm{s}}\bigr)$ 
& $3.058 \pm 0.014$ 
& $3.066_{-0.015}^{+0.013}$
& $3.067_{-0.015}^{+0.013}$
& $3.066 \pm 0.014$ \\

$\Delta$ 
& $0.46 \pm 0.25$ 
& $0.01 \pm 0.16$
& $-0.05 \pm 0.15$
& $0.03 \pm 0.19$ \\

\hline

$H_0 (\mathrm{km\,s^{-1}\,Mpc^{-1}})$ 
& $69.79 \pm 0.89$ 
& $68.12 \pm 0.59$
& $67.91 \pm 0.56$
& $68.20 \pm 0.73$ \\

$\sigma_{8}$ 
& $0.833 \pm 0.011$ 
& $0.8153 \pm 0.0080$
& $0.8131 \pm 0.0081$
& $0.8162 \pm 0.0093$ \\

$S_{8}$ 
& $0.8204 \pm 0.0070$ 
& $0.8210 \pm 0.0070$
& $0.8210 \pm 0.0071$
& $0.8211 \pm 0.0072$ \\

$\Omega_{\mathrm{m}}$ 
& $0.2914 \pm 0.0070$ 
& $0.3042 \pm 0.0052$
& $0.3060 \pm 0.0050$
& $0.3037 \pm 0.0062$ \\

$r_d\text{(Mpc)}$ 
& $147.38 \pm 0.19$ 
& $147.54 \pm 0.19$
& $147.56 \pm 0.19$
& $147.53 \pm 0.19$ \\

\hline

$\Delta{\chi^{2}_{\text{MAP}}}$
& -0.37 
& +0.33
& -2.00 
& -0.11 \\

$N\sigma$ 
& 0.61  
& 0.00
& 1.41 
& 0.33 \\

$\ln B_{i,j}$
& 0.65
& 0.14
& 0.24
& 0.58 \\

\hline
\end{tabular}
}
\caption{This table shows the constraints on the GEDE parameters, presenting the mean values and their uncertainties at the 68\% ($1\sigma$) confidence level, obtained using DESI DR2 combined with CMB and SNe~Ia datasets (Pantheon$^+$, DES-Dovekie, and Union3).}\label{tab_2}
\end{table*}

\begin{figure*}
\centering
\includegraphics[scale=0.44]{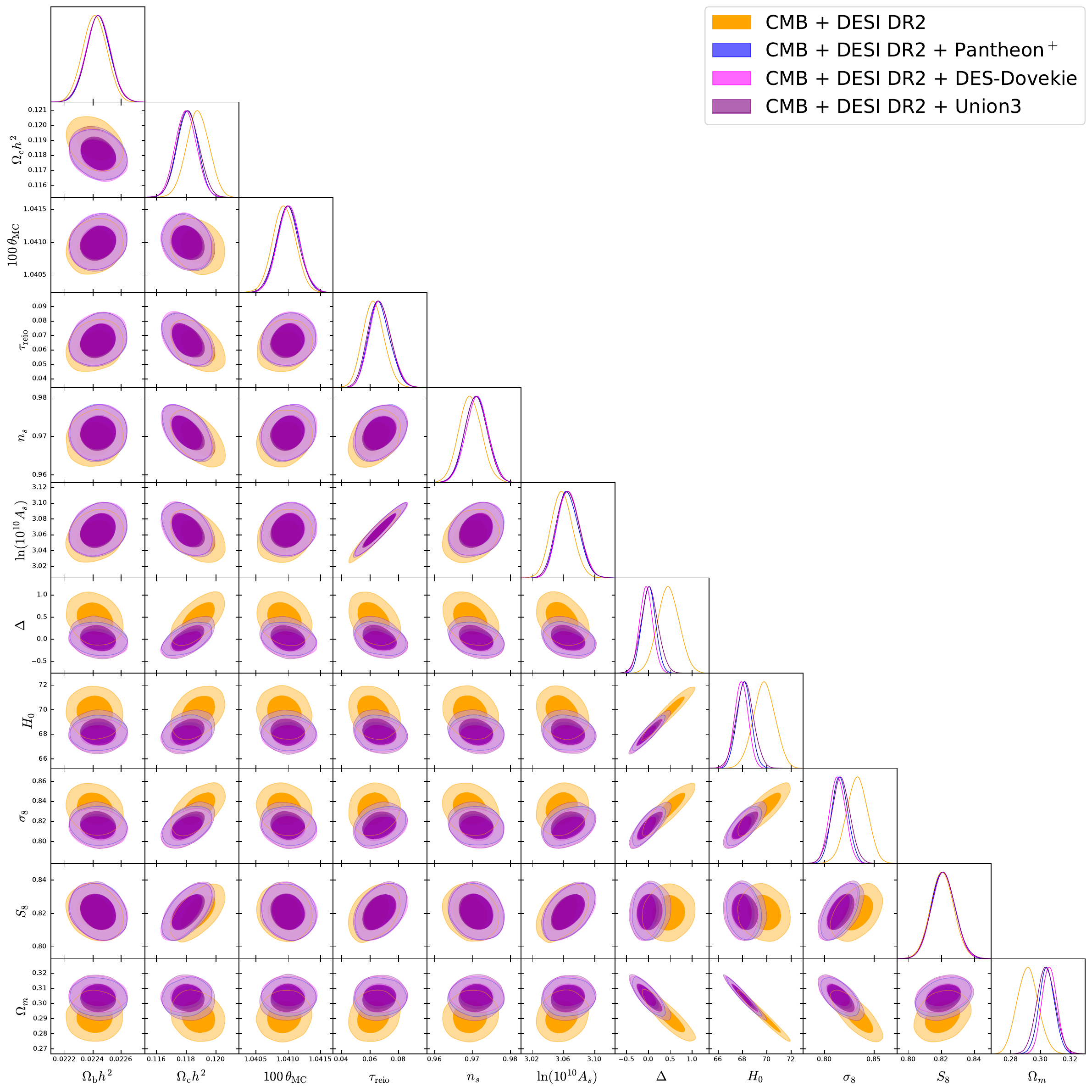}
\caption{The figure shows the corner plot of the GEDE model at 68\% ($1\sigma$) and 95\% ($2\sigma$) confidence levels using DESI DR2 combined with CMB and SNe~Ia datasets (Pantheon$^+$, DES-Dovekie, and Union3), shown as superimposed contours for the different dataset combinations.}\label{fig_1}
\end{figure*}

\begin{figure}
\centering
\includegraphics[scale=0.55]{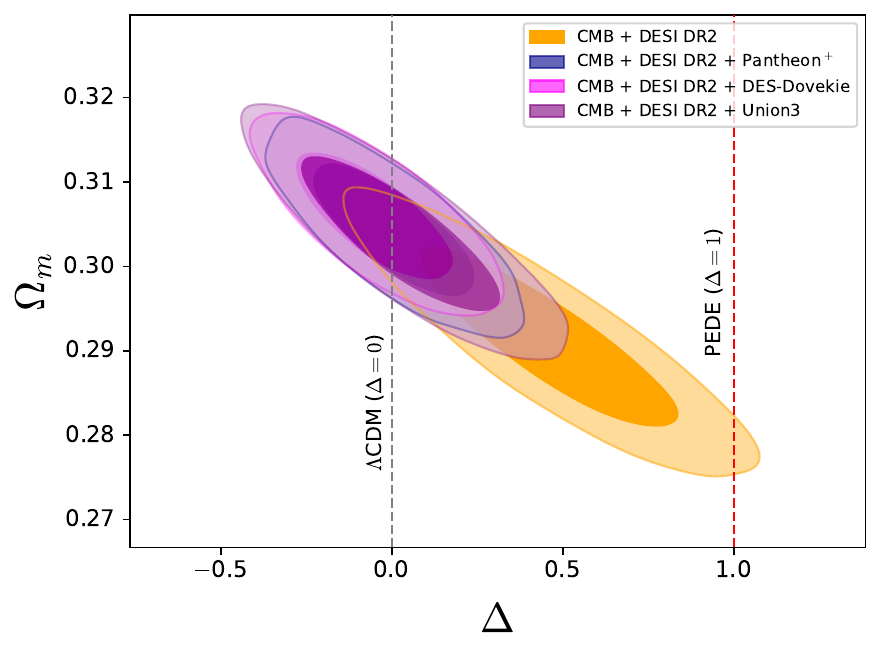}
\caption{The figure shows the 2D marginalized confidence contours of the $\{\Delta-\Omega_m\}$ plane of the GEDE model at 68\% ($1\sigma$) and 95\% ($2\sigma$) confidence levels using DESI DR2 combined with CMB and SNe~Ia datasets (Pantheon$^+$, DES-Dovekie, and Union3), shown as superimposed contours for the different dataset combinations.}\label{fig_2}
\end{figure}

\begin{figure*}
\centering
\includegraphics[scale=0.42]{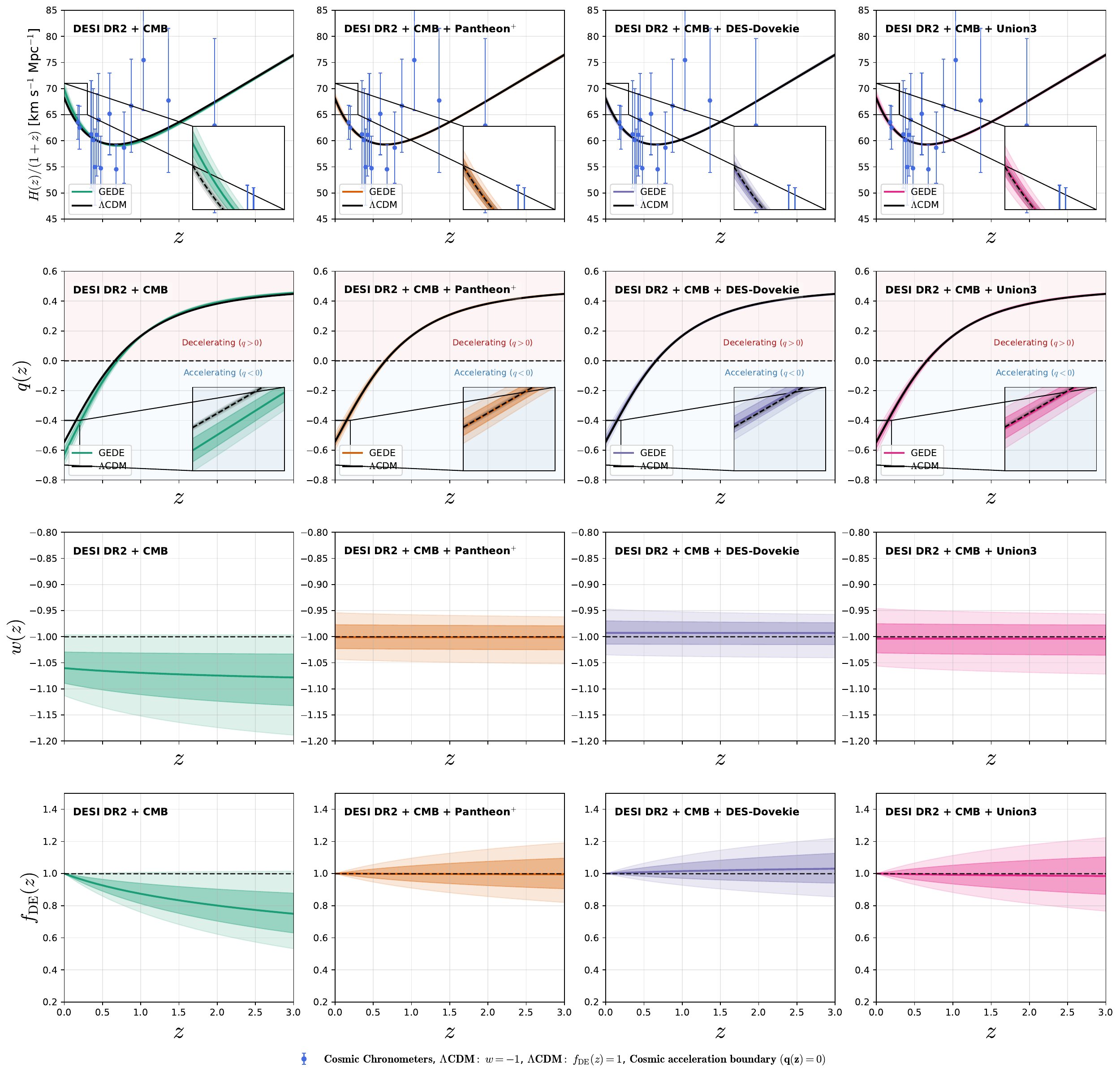}
\caption{The figure shows the evolution of $H(z)/(1+z)$ (1st row), $q(z)$ (2nd row), $w(z)$ (3rd row), and $f_{\mathrm{DE}}(z)$ (4th row) for the GEDE and $\Lambda$CDM models as functions of $z$. The solid lines represent the mean predictions, while the light and dark shaded regions correspond to the $1\sigma$ and $2\sigma$ confidence intervals, respectively.}\label{fig_3}
\end{figure*}

\section{Results and Conclusions}\label{sec_4}
In Fig.~\ref{fig_1}, we show the contour plots for the GEDE model parameters, showing the 68\% ($1\sigma$) and 95\% ($2\sigma$) confidence regions obtained from the combination of DESI DR2, CMB, and SNe~Ia samples (Pantheon$+$, DES-Dovekie, and Union3). The off-diagonal panels show the 2-dimensional (2D) posterior distributions for each pair of parameters, where the inner and outer contours correspond to the 68\% ($1\sigma$) and 95\% ($2\sigma$) confidence levels, respectively. The diagonal panels show the 1D marginalized posterior distributions of the GEDE cosmological parameters. Table~\ref{tab_2} shows the mean values and their uncertainties at the 68\% ($1\sigma$) confidence level, obtained using DESI DR2 combined with CMB and SNe~Ia datasets (Pantheon$^+$, DES-Dovekie, and Union3).

First, we discuss the $H_0$ tension in the GEDE model after DESI DR2. To address the $H_0$ tension, it is essential to match the value of $H_0$ to that reported by the SH0ES team, which is obtained using the distance ladder method based on Cepheid-calibrated SNe~Ia, yielding $(73.04 \pm 1.04),\mathrm{km,s^{-1},Mpc^{-1}}$~\cite{riess2022comprehensive}. At the same time, this must be achieved in a manner consistent with reducing the sound horizon $r_d$ by approximately 7\% \cite{knox2020hubble}. It is also important to note that reducing the sound horizon $r_d$ by approximately $\sim 7\%$ generally requires a higher physical matter density. However, this typically leads to an increase in the clustering amplitude, thereby exacerbating the $S_8 \equiv \sigma_8 \sqrt{\Omega_m/0.3}$ tension. As a result, any attempt to resolve the $H_0$ tension in this way may worsen the already existing discrepancy between $\Lambda$CDM predictions and weak lensing (WL) observations \cite{vagnozzi2023seven,jedamzik2021reducing}.

In the case of the GEDE model, we do not find any significant alleviation of the $H_0$ value across the different dataset combinations. As shown in Table~\ref{tab_2}, the inferred values of $H_0$ remain in the range $H_0 \sim 67.9$–$69.8\ \mathrm{km,s^{-1},Mpc^{-1}}$, which is still in clear tension with the SH0ES measurement. At the same time, the sound horizon remains nearly unchanged, with $r_d \simeq 147.4$–$147.6\ \mathrm{Mpc}$ for all dataset combinations. This indicates that the GEDE model does not provide the required reduction in the sound horizon, $r_d$, of order $\sim 7\%$. Indeed, although the GEDE model introduces a purely post-recombination modification to the background expansion history, it leaves the physics of the early Universe completely unchanged. Consequently, it cannot reduce the sound horizon at recombination and therefore remains unable to resolve the $H_0$ tension, as thoroughly detailed in the literature \cite{Bernal:2016gxb,Aylor:2018drw,Knox:2019rjx,Efstathiou:2021ocp,Jiang:2024xnu}.

For the GEDE model, we obtain $\sigma_8 \sim 0.813\text{-}0.833$ and $S_8 \sim 0.8204\text{--}0.8211$ across the different dataset combinations. These values remain consistently high and are in close agreement with the $\Lambda$CDM predictions from CMB measurements. Consequently, the GEDE model does not lead to any significant reduction in $S_8$, and therefore does not alleviate the $S_8$ tension with weak lensing observations, which typically prefer lower values of $S_8$ \cite{abbott2022dark,asgari2021kids}. This suggests that, similar to the $H_0$ tension, the GEDE model fails to provide a solution to both tensions in the light of cosmological observations.

In Fig.~\ref{fig_2}, we show the 2D marginalized confidence contours at 68\% ($1\sigma$) and 95\% ($2\sigma$) confidence levels in the $\{\Delta-\Omega_m\}$ plane for the GEDE model. For the CMB + DESI DR2 dataset combination, we find a preferred value of $\Delta = 0.46 \pm 0.25$, which deviates from the $\Lambda$CDM reference value of $\Delta = 0$ at the $\sim 1.8\sigma$ level. When the Pantheon$^+$ sample is combined with CMB and DESI DR2, the GEDE model yields $\Delta = 0.01 \pm 0.16$, corresponding to a negligible deviation of $0.06\sigma$ from the $\Lambda$CDM reference value $\Delta = 0$. Similarly, for the CMB + DESI DR2 + DES-Dovekie combination, we obtain $\Delta = -0.05 \pm 0.15$, corresponding to a deviation of $0.33\sigma$. For the CMB + DESI DR2 + Union3 dataset, we find $\Delta = 0.03 \pm 0.19$, corresponding to a deviation of $0.16\sigma$. These results indicate that, once SNe~Ia data are included, the parameter $\Delta$ becomes fully consistent with $\Lambda$CDM, showing no statistically significant deviation. This behavior is expected, as Type Ia supernova observations place very stringent constraints on emergent dark energy models and, more generally, on late-time modifications to the expansion history of the Universe \cite{Pedrotti:2025ccw}.

Fig.~\ref{fig_3} shows the evolution of $H(z)/(1+z)$ (1st row), $q(z)$ (2nd row), $w(z)$ (3rd row), and $f_{\mathrm{DE}}(z)$ (4th row) for the GEDE and $\Lambda$CDM models as functions of $z$. The solid lines represent the mean predictions, while the light and dark shaded regions correspond to the $1\sigma$ and $2\sigma$ confidence intervals, respectively. In the first row, we show the evolution of $H(z)/(1+z)$ for both the $\Lambda$CDM and GEDE models. It can be observed that, in the case of the DESI DR2 + CMB combination, the predicted mean value of the GEDE model is slightly higher than that of $\Lambda$CDM. However, when Pantheon$^+$, DES-Dovekie, and Union3 datasets are included, the mean values of both models become very close to each other. This behavior can be clearly seen in the zoomed-in panels of each plot. The results are also compared with cosmic chronometer (CC) measurements from \cite{moresco2012improved,moresco2015raising,moresco20166}. In the second row, we show the evolution of the deceleration parameter $q(z)$ for the $\Lambda$CDM and GEDE models. It can be observed that both models shows a smooth transition from a decelerating to an accelerating phase. In the case of the DESI DR2 + CMB combination, the GEDE model predicts a slightly higher acceleration compared to the $\Lambda$CDM model. However, when Pantheon$^+$, DES-Dovekie, and Union3 datasets are included, the present-day values of the deceleration parameter ($q_0$) for both models become nearly identical.

In the third row, we show the evolution of the $w(z)$ parameter for the GEDE model, while $\Lambda$CDM assumes a constant value $w = -1$. In the case of the CMB + DESI DR2 combination, the mean value of $w(z)$ remains below $-1$. Indeed, the GEDE model shows a full phantom regime ($w < -1$ at all epochs), consistent with $w_0 < -1$ and $w_0 + w_a < -1$~\cite{ratra1988cosmological}, a similar behavior can be observed when we combine Union3 with CMB + DESI DR2. When we combine DES-Dovekie with CMB + DESI DR2, the mean predictions of the GEDE model show a full quintessence behavior ($w > -1$ at all epochs, thus $w_0 > -1$ and $w_0 + w_a > -1$)~\cite{caldwell2002phantom}. Interestingly, when we combine Pantheon$^+$ with DESI DR2 + CMB, the mean value becomes close to the $\Lambda$CDM prediction, $w = -1$. The fourth column shows the evolution of $f_{\mathrm{DE}}(z)$ as a function of redshift. Indeed, for all four combinations, at $z = 0$, $f_{\mathrm{DE}}(z)$ converges to $f_{\mathrm{DE}}(z) = 1$.

The corresponding Gaussian significance levels are $N\sigma = 0.61$, $0.00$, $1.41$, and $0.33$, respectively, which remain below the $2\sigma$ threshold. This implies that the emergence of emergent dark energy is not statistically significant for any of the dataset combinations considered. The Bayesian model comparison is performed using the natural logarithm of the Bayes factor, $\ln B_{i,j}$, interpreted according to the Jeffreys scale. For the dataset combinations considered, we obtain $\ln B_{i,j} = 0.65$, $0.14$, $0.24$, and $0.58$, which all fall within the regime of inconclusive evidence ($\ln B_{i,j} < 1$). This indicates that there is no statistically significant preference for the GEDE model over the standard $\Lambda$CDM scenario.

In this work, we revisited the phenomenology of the GEDE model using the DESI DR2 and joint ACT, SPT, and Planck data, along with SNe Ia datasets (Pantheon$^+$, DES-Dovekie, and Union3). Overall, our analysis shows that the GEDE model remains fully consistent with the standard $\Lambda$CDM cosmology, with no statistically significant preference for deviations when all datasets are combined. In particular, once SNe Ia data are included, the key model parameter $\Delta$ becomes entirely consistent with the $\Lambda$CDM reference value $\Delta = 0$, with deviations well below the $2\sigma$ level.

Despite this overall consistency, the GEDE model does not show the phantom crossing predicted by the DESI DR2 results. However, from the $w(z)$ evolution, one can see that the GEDE model can be effectively described by a single scalar field, behaving either as a full phantom regime ($w < -1$) or as quintessence ($w > -1$), depending on the dataset combination. Interestingly, when combining DESI DR2 + CMB + Pantheon$^+$, the model converges closely to the $\Lambda$CDM prediction, with $w \simeq -1$. This behavior is in contrast with the DESI DR2 predictions, which favor a Quintom-B–type dark energy scenario ($w = -1$, i.e., $w < -1$ in the past but $w > -1$ today, corresponding to $w_0 > -1$ and $w_0 + w_a < -1$) \cite{gu2025dynamical}. Therefore, our results show that, although the GEDE model remains consistent with $\Lambda$CDM, it does not agree with the type of dynamical dark energy behavior suggested by DESI DR2.

\section*{Acknowledgements}
S.C.  acknowledges the Istituto Italiano di Fisica Nucleare (INFN) iniziative specifiche QGSKY and MOONLIGHT2 and  the Gruppo Nazionale di Fisica Matematica (GNFM)  of Istituto Nazionale di Alta Matematica (INDAM) for the support. This paper is based upon work from COST Action CA21136 -- Addressing observational tensions in cosmology with systematics and fundamental physics (CosmoVerse), supported by COST (European Cooperation in Science and Technology). Himanshu Chaudhary is very thankful to Prof. Gao Xianlong from the Department of Physics, Zhejiang Normal University,
for his kind support and help during this research.

\bibliographystyle{elsarticle-num}
\bibliography{mybib.bib}

\end{document}